\documentclass[10pt,letterpaper]{article}
\usepackage{opex3}
\usepackage{color}
\usepackage{cite}
\usepackage{amsmath}

\begin{document}

\title{Evanescent field trapping of nanoparticles using nanostructured ultrathin optical fibers}

\author{Mark Daly,$^1$ Viet Giang Truong,$^1$ and S\' ile Nic Chormaic$^{1}$}

\address{$^1$Light-Matter Interactions Unit, OIST Graduate University, Onna, Okinawa, 904-0495, Japan}

\email{$^*$sile.nicchormaic@oist.jp} %% email address is required



\begin{abstract}
While conventional optical trapping techniques can trap objects with submicron dimensions, the underlying limits imposed by the diffraction of light generally restrict their use to larger or higher refractive index particles. As the index and diameter decrease, the trapping difficulty rapidly increases; hence, the power requirements for stable trapping become so large as to quickly denature the trapped objects in such diffraction-limited systems. Here, we present an evanescent field-based device capable of confining low index nanoscale particles using modest optical powers as low as 1.2 mW, with additional applications in the field of cold atom trapping. Our experiment uses a nanostructured optical micro-nanofiber to trap 200 nm, low index contrast, fluorescent particles within the structured region, thereby overcoming diffraction limitations. We analyze the trapping potential of this device both experimentally and theoretically, and show how strong optical traps are achieved with low input powers.
\end{abstract}

\ocis{(140.7010) Laser trapping; (060.2310) Fiber optics; (220.4241) Nanostructure fabrication.} % REPLACE WITH CORRECT OCIS CODES FOR YOUR ARTICLE, MINIMUM OF TWO; Avoid using the OCIS codes for “General” or “General science” whenever possible.

%%%%%%%%%%%%%%%%%%%%%%% References %%%%%%%%%%%%%%%%%%%%%%%%%

%%%%%%%%%%%%%%%%%%%%%%%%%%  body  %%%%%%%%%%%%%%%%%%%%%%%%%%

\section{Introduction}

The concept that light could impart forces to matter can be traced back to early suggestions by Kepler in the 1600s. He observed that the tail of a comet seemed to incongruously point in a direction retrograde to its motion \cite{Kepler}. The tail was seen to point away from the sun and hence the notion of radiation pressure was born. With the advent of coherent light sources, many groups began to make heavy use of optical forces experimentally.  For example, Doppler cooling, the precursor to many laser cooling techniques, gave way to the field of atom trapping \cite{Ashkin78}. Ashkin \textemdash  often considered to be a pioneer in the field of optical trapping\textemdash  proposed that Gaussian beams could be used to trap silica microparticles using a technique that would later become known as optical tweezing \cite{Ashkin86}. At this early stage, it was apparent that optical tweezers had fundamental operational constraints due to the diffraction-limited spot size of the trapping beam. Today, nano-optical techniques, such as near-field optics and plasmonics, provide primary solutions to this problem \cite{Marago13,Daly15}. Photonic crystal cavities \cite{Song05,Jing16}, plasmonic double nano-holes \cite{Zehtabi12,Kotnala14}, slot waveguides \cite{Anderson06,soltani14}, and micro-nanofibers \cite{Tong03,Hoffman14} are just some of the devices which can  confine light locally to  regions smaller than achievable using diffraction-limited systems. Aside from modifying how the trapping fields are generated, it is also possible to change the material of the particles to be trapped, thereby reducing the difficulties associated with trapping submicron particles. For example, higher index particles, such as gold nanoparticles \cite{Ploschner12,Brzobohaty15} are excellent  candidates for nanoscale trapping but have associated problems with heat generation. Other high index particles such as nanodiamonds \cite{Geiselmann13} and Titania particles \cite{Jannasch12} are also easier to trap; however, biologically-relevant materials typically have low refractive indices, thereby negating the trapping advantages associated with higher index particles.  

In this article, we discuss a nanostructured, evanescent optical trapping device based on the combination of a slot waveguide with a micro-nanofiber (MNF) \cite{daly2014nanostructured,daly15ot,daly15fab}. MNFs are extremely versatile due to their compact size, enabling them to be integrated noninvasively into many systems, such as optical tweezers \cite{xin2011targeted,maimaiti2015higher,Gusachenko15} and cold atom clouds \cite{nieddu15optical}.  When light propagates through an MNF, a significant portion of the electric field exists outside the waveguide as an evanescent component, allowing for easy interaction between the guided light field and the surrounding medium. We work with a nanostructured MNF with an overall waist of 1.4 $\mu$m guiding light with a wavelength of 980 nm. Previously, unmodified MNFs have been used for various experimental configurations such as (i) optical trapping of dielectric particles \cite{Skelton12,Li13,Gana14,Frawley14,Cheng16}, (ii)  cavity quantum electrodynamics (cQED) using single quantum emitters \cite{yalla2014cavity}, (iii) light coupling in and out of whispering gallery resonators \cite{knight1997phase}, and (iv) trapping and probing cold atomic systems \cite{Mitsch14,Beguin14,Kato15,Kumar15} or atomic vapors \cite{watkins2013observation,jones2015ladder}. More recently, MNFs have been modified  to increase their versatility across a range of fields, through, for example, the incorporation of SNOM tips \cite{Xin12} or extraordinary transmission apertures \cite{neumann2010extraordinary}. Evanescent field trapping has also been realised using what is known as a slot waveguide, as first demonstrated by Yang et al. \cite{yang2009optical} who trapped 75 nm dielectric nanoparticles. Laser powers of 250-300 mW provided stable trapping against a constant fluid flow. The high refractive index contrast between the Si slot and the surrounding water, along with the small slot separation ($<$100 nm), produced a quasi-TE mode with a large field discontinuity across the boundary that was used for trapping.

While other trapping techniques, such as self-induced back action (SIBA) \cite{juan2009self}, can confine particles in three dimensions with low optical powers, they lack adequate control over the particle's position. For optical fibers, dynamic three dimensional control over the position of trapped particles becomes difficult, but recent developments using orbital angular momentum carrying beams for particle trapping \cite{jones2015optical} may soon allow for the spatial translation of particles, whilst still maintaining strong trap stiffnesses. 

\begin{figure}[!h]
\centering \includegraphics[width=0.5\columnwidth]{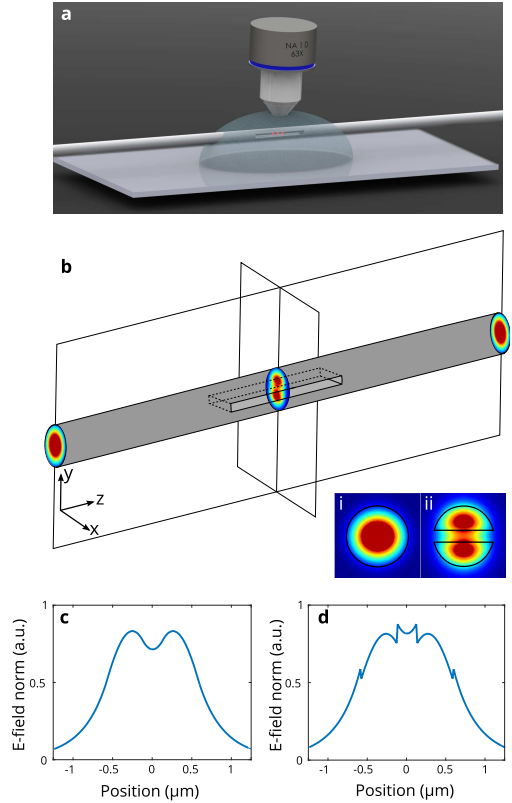}
\caption{\label{fig:fig1}(a): Representation of the slotted tapered optical fiber (STOF) in a solution of red fluorescent silica nanoparticles. A 63x immersion lens is used to image the system. (b): A schematic showing the STOF section of the optical fiber with the fundamental fiber mode (i) seen at either side of the cavity region and the fundamental STOF mode (ii) at the center. (c) and (d) show typical electric field norm along a line cutting through the origin along y for polarizations parallel to and perpendicular to the slot wall, respectively. The field within the slot can be up to 1.7 times higher than the field at the outer fiber surfaces although variations in the STOF dimensions can drastically alter this. The origin is taken to be at the center of the slot.}
\end{figure}

The slotted tapered optical fiber (STOF) used in this work is a device which exploits the overlapping evanescent fields  of a slot waveguide-like structure to further enhance its trapping ability, both for atoms \cite{daly2014nanostructured} and submicron particle trapping. We sought to create an entirely fiber-based trapping device using evanescent fields to localize particles with a high degree of control in regimes inaccessible to standard MNFs. A nanoscale slot is introduced at the waist of an MNF using focused ion beam (FIB) milling, thereby creating a slot waveguide-like region where the optical forces are greatly enhanced. This provides three dimensional confinement within a small trapping volume, while also providing the potential for one dimensional position control of the trapped particles along the slot through the use of a sliding standing wave, i.e. a particle conveyor belt \cite{Cizmar06}. Here, we demonstrate trapping of 200 nm silica particles using a STOF. We also show how light transmitted through the fiber pigtails either side of the STOF can be analyzed to determine the trap characteristics.  Finite element method (FEM) and finite difference time domain (FDTD) simulations are used to simulate the STOF modes and these are subsequently used to model the forces acting on the particles using perturbative and energy density methods.

\section{Experimental Setup}
\subsection*{Micro-nanofibers and slotted tapered optical fibers}

Commercial optical fibers guide light in what is known as the weakly-guided regime, wherein the refractive index contrast between the core of the fiber and the surrounding cladding is quite low ($n_{core}-n_{cladding}<0.01$).  Most of the light is contained within an area defined by the mode field diameter, which is much smaller than the fiber's total cross-sectional area. When an optical fiber is tapered over a heat source such that its diameter is close to, or below, the wavelength of the guided light, the distinction between the core and cladding region is no longer valid. The surrounding medium (in our case, water) becomes the new cladding and the original cladding is now viewed as the core material. These MNFs  operate in the strongly guided regime as the refractive index contrast becomes high. The evanescent fields produced in MNFs extend far \textemdash when compared to the waveguide dimensions \textemdash into the surrounding medium and can  interact with particles located at several 100s of nm from the fiber's surface. We work with optical fibers which have been tapered using a heat-and-pull method so that the waist diameter is typically of the order of the wavelength \cite{ward2014contributed}.We used a hydrogen-oxygen flame mixed in a 2:1 ratio to provide a clean-burning source. The untapered optical fiber is stripped of its outer acrylic layer and clamped to the stages then placed into the flame. The stages pull both sides of the fiber away from the flame, causing the fiber, which is now in a molten state, to taper. By controlling the speed of the stages, the length of the pull, and the flame size, MNFs with specified diameters can be produced. The MNFs were fabricated from single-mode optical fiber in the 980\textendash1600 nm regime. Slotted tapered optical fibers, or STOFs, are nanostructured MNFs which have had a section of their waist removed. The slotted tapered optical fibers were created using a three step process which involved the initial MNF fabrication process using standard heat-and-pull techniques, an indium tin oxide (ITO) sputter coating process, and finally a focused ion beam milling process to introduce a slot to the MNF. A 5 nm layer of ITO is necessary to provide sufficient charge mitigation at the MNF surface as an uncoated MNF would be subject to large dielectric charging effects during the FIB process, thereby making the subsequent etching of submicron features impossible. This new technique enables us to 'write' high resolution structures directly onto the MNF in a transmission-preserving, three-step process. The STOF geometry is illustrated  in Fig. \ref{fig:fig1}.

\begin{figure*}[!ht]
\centering 
\includegraphics[width=0.75\linewidth]{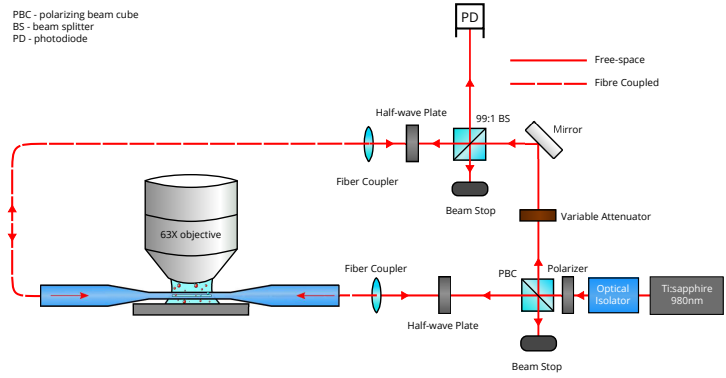}
\caption{\label{fig:fig2} Optical setup used to trap nanoparticles. 980 nm light from a Ti:sapphire laser is passed through a polarizing beam splitter to split the beam while providing some initial control over the power balance. From here the beams are passed through polarization control optics and finally fiber coupled to the STOF. Transmission data is collected via a photodiode}
\end{figure*}

\subsection*{Field distribution}

As expected, the electric fields of a STOF differ greatly from those of a typical tapered fiber. Due to the physical asymmetry introduced to the MNF at the slot region, the polarization of the guided light plays a larger role in the field distribution. To maximize the field at the slot, the polarization must  be perpendicular to the walls of the slotted region. This removes the continuity requirement of the electric field at the dielectric boundary, thereby allowing a large portion of the field to exist outside of the waveguide (see Fig.\ref{fig:fig1}(d)). The strength of the field at the slotted region depends on a number of parameters: the diameter, the slot width, the polarization and the wavelength of the guided light. We chose a 1.4 $\mu$m fiber waist with a slot width of 300 nm. In general, the field strength increases with decreasing slot size, but, since we have an additional requirement that the STOF opening must be large enough to facilitate the entry of submicron particles, the slot width used was the minimal possible while still being practical for particle trapping.  When the slot region is excited by the fundamental mode of an MNF, as illustrated in Fig. \ref{fig:fig1}(b) inset (i), a 'fundamental' type mode  is excited in either section, see Fig. \ref{fig:fig1}(b) inset (ii). Slot waveguides can exhibit symmetric or anti-symmetric modes depending on the phase difference between the upper and lower sections. The fundamental mode of an MNF has an approximately uniform phase front, so we neglect the possibility of anti-symmetric modes in the following discussion.

\subsection*{Experimental outline}
Optical trapping forces are typically divided into two categories in optical trapping experiments. In the dipole approximation, where $[n_{p}/n_{m}]k a \ll 1$ where $k$ is the wave number (i.e. $2\pi / \lambda$), $a$ is the radius of the particle, and $n_p$ and $n_m$ are the refractive indices of the particle and medium, respectively, the force can be decomposed into the gradient force ($\textbf{F}_{g}\propto \frac{1}{2}\alpha \nabla \textbf{E}^2$) and the scattering force ($\textbf{F}_{s}\propto I(\textbf{r})\hat{z}$) \cite{jones2015optical}, where $\alpha$ is the real component of the polarizability of the particle, \textbf{E} is the electric field, and \textit{I}(\textbf{r}) is the optical intensity. This formalism is not necessarily accurate for all particle sizes, but gives a qualitative and intuitive picture of how the local electric fields affect particles placed within them. 

For the case of unidirectional excitation of a STOF waveguide, the gradient force, which seeks to pull particles towards regions of high intensity,  draws particles towards the walls at the center of the slot, while the scattering force  propels particles in the direction of propagation of the trapping laser field. To produce a trap with longitudinal confinement, a standing wave is necessary. This provides an extra degree of confinement for the particles, as well as increasing the overall trap efficiency due to the cancellation of the scattering force components, thereby improving the axial trapping strength. A more in-depth analysis of the trap is made using a combination of FDTD/FEM models and various optical trapping models. 

We introduced a low density nanoparticle solution between the water immersion 63x lens and the STOF, shown in Fig. \ref{fig:fig2}. The low density solution was used to prevent large numbers of particles occluding the slot.  We used a particle solution of approximately $10^{9}$ particles/ml, equivalent to an average particle occupancy of $<$ 1 over the volume of the slot. Fluorescent nanoparticles were used to increase visibility of the system and we collected data visually, using a high sensitivity, fluorescence camera.  Transmission and fluorescence data were also collected through the fiber using either a photodiode (for transmission) or a single photon counting module (for fluorescence).  

\begin{figure*}[!ht]
\centering 
\includegraphics[width=0.65\linewidth]{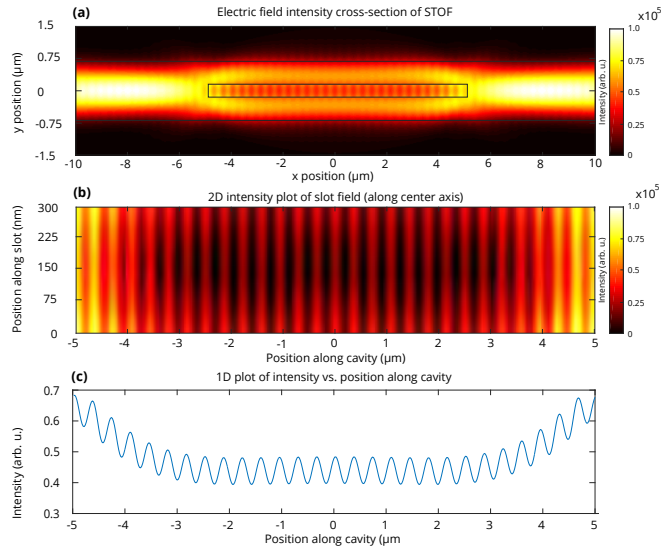}
\caption{\label{fig:fig3} (a): Results of FDTD analysis showing a cross-section of the STOF. The mode evolves from the fundamental mode of the MNF to the fundamental mode of the STOF at the center and back to the fundamental mode of the MNF with little loss. (b): Electric field intensity within the 10 $\mu$m $\times$ 300 nm slot in a 1.4 $\mu$m diameter MNF. The field increases in strength near the slot walls. (c): 1D plot of the electric field across the center of the STOF to show the variation in the field as a function of the distance along the cavity. The field stabilizes at the center of the cavity.}
\end{figure*}

\subsection*{Polarization preparation}
Polarization preparation of light in an MNF is often a source of contention as temperature and pressure variations (i.e. stresses and strains) along the fiber can cause the polarization to rotate, leading to little correlation between the input and output polarization states. Because of this behavior, polarization measurement methods, such as the observation of Rayleigh scattering along the fiber waist, need to be used \cite{szczurowski2011differential}. In our experiment, we counter-propagated 980 nm  light from a Ti:Sapphire laser through the STOF to provide a 3-dimensional trapping potential. The standing wave pattern that was set up within the wave guiding structure extended to the evanescent field, which then interacted with nearby particles to produce a periodic potential within the cavity section of the STOF. Two half-wave plates placed in the path of the two beams gave us fine control over the input polarization states. By monitoring the output at opposite ends of the STOF, we obtained an estimate for the polarization at the slot. However, by directly monitoring the slot region using a 63x water immersion objective lens, we were better able to determine the polarization state at the STOF region via the intensity of the scattered light. When the light was polarized perpendicular to the plane of the slot i.e. along the \textit{y}-axis, scattering was further enhanced due to the increased fraction of light contained within the small region. The slot is viewed in the \textit{yz}-plane where the full slot opening is observed to ensure that the maximum scattering is at the correct polarization state.

\section{Numerical Analysis}
A thorough calculation of the electric fields, using both FDTD and FEM methods, provided us with reliable estimates for the optical forces in the STOF system. The optical fields of the device were calculated both in  the presence of, and without, the particles to be trapped. Optical forces on small particles are often described using the dipole approximation, where the size of the particle must  be much less than the wavelength of the trapping beams. For 200 nm particles this criterion is not quite satisfied. To this end, we sought to make a comparison between the dipole gradient and scattering forces, and the more standard methods of force calculation for medium-sized particles. The total optical force in mid-sized optical trapping systems, $\textbf{F}_{MST}$, is often calculated using a surface integral of the dot product of the Maxwell stress tensor, \textbf{T}, with the surface normal, \textbf{n}, such that 
\begin{equation}
\textbf{F}_{MST}=\oint_S(\textbf{T}.\textbf{n})da.
\end{equation}

\begin{figure}[!h]
\centering 
\includegraphics[width=0.4\linewidth]{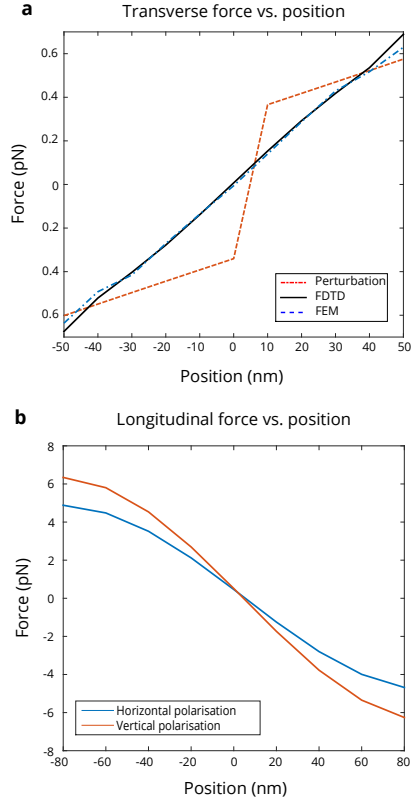}
\caption{\label{fig:fig4} (a): Forces on a 200 nm particle moving perpendicularly between the upper and lower walls of a STOF as determined using optical fields from FDTD and FEM calculations and Eqn. 2, compared to a perturbative approach using the optical fields of the cavity in the absence of a particle as modeled using the FEM. 1 W of power was used in all simulations. (b): Longitudinal trapping force for two orthogonal polarization states showing the increased trapping forces for the vertical polarization state.}
\end{figure}

\noindent Here, $da$ is the unit area element.
While this is an accurate method, it can be somewhat difficult to implement when the boundary of the system is ill-defined due to the mesh shape and/or the step size of the FEM/FDTD method used. As an alternative, we used an equivalent form derived from the Minkowski formalism for calculating the force; this method relies on the gradient of the electric permittivity, a value which can be easily extracted from the optical force calculation \cite{brevik1979experiments}, such that the force is given by

\begin{equation}
\textbf{F}_{min}=-\frac{1}{4} \epsilon_0 \iiint _V\textbf{E}.\textbf{E}\nabla\epsilon_rdV,
\end{equation}

\noindent where $\epsilon_0$ is the permittivity of free space, $\epsilon_r$ is the relative permittivity, \textbf{E} is the electric field and $dV$ is the unit volume element. Accepted values of the refractive indices of silica and polystyrene at a wavelength of 980 nm were used in all calculations. The effect of the 5 nm layer of indium tin oxide was ignored due to its negligible influence on the MNF modes. Force measurements using data from the FDTD and FEM methods give almost identical results for the trapping forces within the slot except near the slot walls where a maximum discrepancy of 7.6\% was found. We assume this discrepancy to be associated with the dynamic meshing of the FEM. To achieve similar resolution near the boundaries of the slot walls in an FDTD calculation, a significant increase in computation time would be required. The dipole approximation for the force calculation proved unreliable in regions where the local gradient was insignificant, but became more accurate as this gradient increased. This method could be improved  by considering the particle as a distributed dipole, but this treatment is beyond the scope of this work. FDTD images of the optical fields of the STOF are given in Fig. \ref{fig:fig3}, and a comparison of the trapping forces for different particle locations inside the slot are given in Fig. \ref{fig:fig4}. Considering the close agreement between the FDTD and FEM simulations, we chose to largely model the system using the FDTD method due to the reduced memory requirements and regular grid pattern. The dipole approximation,

\begin{equation}
F_{dipole}=\frac{1}{2}\alpha\nabla\textbf{E}^2,
\end{equation}

\noindent was also used to provide a contrast to the force calculation using the Minkowski formalism. The nature of Mie scattering requires smaller particle dimensions before one can neglect the higher order poles in the multipole expansion, hence the discrepancy between the two methods.

At the ends of the slot, the local electric fields were found to increase due to reflections off the walls. Proposed solutions to this problem involve the introduction of a slot with tapered ends to allow the modes to evolve adiabatically between the MNF and STOF modes. After $\approx$2 $\mu$m the field within the cavity stabilizes, see Fig. \ref{fig:fig3}(c).
The trapping forces along the directions which run parallel to the STOF wall (along the \textit{x}- and \textit{z}-axes) are considered to have normal restoring forces, but the trap which runs perpendicularly (along the \textit{y}-axis) between the upper and lower walls of the STOF does not experience a standard restoring force.  Here, the optical forces  seek to pull particles towards the wall, at which point any restoring character is lost. Instead, we determined the gradient of the force in this direction, indicated by ${g}_{y}$ in Table \ref{tab:trapstiff}.

\section{Results and trap analysis}

\begin{figure}[!h]
\centering 
\includegraphics[width=\linewidth]{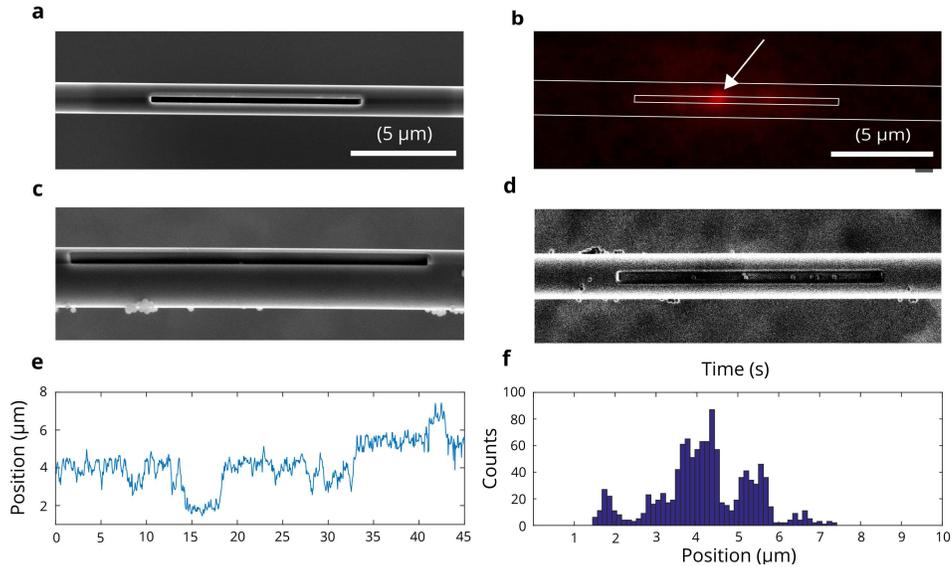}
\caption{\label{fig:fig5}  (a) SEM image of a STOF. (b): Microscope image of a trapped fluorescent particle with an outline of the STOF for clarity \textcolor{blue}{(see visualization 1 \& 2)}. (c) and (d) show SEM images of the fiber after the experiment was performed. Particles can be seen inside the slot was well as on the surface. (e): Particle position versus time along the \textit{z}-axis of the STOF. The particle is seen to spend most of its time near the slot center. Each pixel was found to correspond to a 100 nm $\times$ 100 nm area and Gaussian fits to the particle center enable high resolution tracking. (f): Histogram of the particle positions given in (e) showing  bunching  at regular intervals.}
\end{figure}

The small dimensions of the STOF make it hard to image the slot adequately without resorting to SEM imaging; sample images shown in figures \ref{fig:fig5}(a-d). This problem also applies to the much smaller dimensions of the particles we wish to trap. Fluorescence imaging Fig. \ref{fig:fig5}(b), however, allows us to capture live video of the particles' motion; due to the low-light levels, exposure times of $\approx$ 70 ms are required to actually image the particles. This limits our ability to perform Fourier analyses of the visual data since the trap operates at relatively low trap frequencies, requiring prohibitively long data collection times \cite{vanderHorst:10}. We can, however, track the particles and bin their positions to observe interesting behavior in their motion Fig. \ref{fig:fig5}(e). With an imaging resolution of 13.3 pixels per micron we were able to track the particles' positions to a high degree of accuracy. Gaussian fitting of the positions showed trapping occurs at regular intervals along the central axis of the STOF. In the bright-field it was difficult to distinguish single particle trapping events from multiple particle trapping, but fluorescent imaging indicates that typically more than one particle is trapped. The dynamics of multiple particle trapping may shed some light on the larger spacing between 'stable' trapping positions. Simulations show that two particles in the  trap have non-negligible interactions over distances of approximately 1 $\mu$m and this may explain the observation of stable trapping positions which are multiples of the approximately 350 nm standing wave separation, $\frac{\lambda}{2n_{eff}}$, where $n_{eff}$ is the effective refractive index, as evident from the histogram in Fig. \ref{fig:fig5}(f). We also took SEM images of the slot following the experiment. The devices were left to dry overnight in an enclosure. The images show particles on the fiber surface as well as inside the slot. This does not prove that any trapping occurred, but it does indicate that particles can diffuse freely into the slot.

The unrestricted motion of Brownian particles leads to a characteristic $\frac{1}{f^2}$ noise spectrum. In contrast, for a trapped particle, the power spectral density (PSD) follows a Lorentzian distribution, $\frac{A}{f^2+f_c^2}$, which is derived from the Langevin equations of motion \cite{berg2004power}. Log-log plots of these data allows one to visually interpret this Lorenztian line-shape as the combination of two regimes which overlap at the corner frequency, $f_c$. Everything beyond $f_c$ can be viewed as the unbound motion of free Brownian particles and  behaves as $\frac{1}{f^2}$, while everything below this represents the restricted motion of the trapped particles \cite{juan2009self}.

\begin{table}[h]
\centering
\caption{\bf Trap 'Stiffnesses' for Varying Input powers as determined from FDTD analysis}
\begin{tabular}{cccc}
\hline
Trap 'Stiffness' & 2 mW & 5mW & 10 mW  \\
\hline
$\tilde{k}_x$ & 28 fN $\mu$m$^{-1}$& 69 fN $\mu$m$^{-1}$ & 138 fN $\mu$m$^{-1}$\\
$k_z$ &202 fN $\mu$m$^{-1}$ & 510 fN $\mu$m$^{-1}$& 1020 fN $\mu$m$^{-1}$ \\
$g_y$ & 102 fN $\mu$m$^{-1}$ & 255 fN $\mu$m$^{-1}$ & 510 fN $\mu$m$^{-1}$\\

\hline
\end{tabular}
  \label{tab:trapstiff}
\end{table}

\begin{figure}[!h]
\centering 
\includegraphics[width=.4\linewidth]{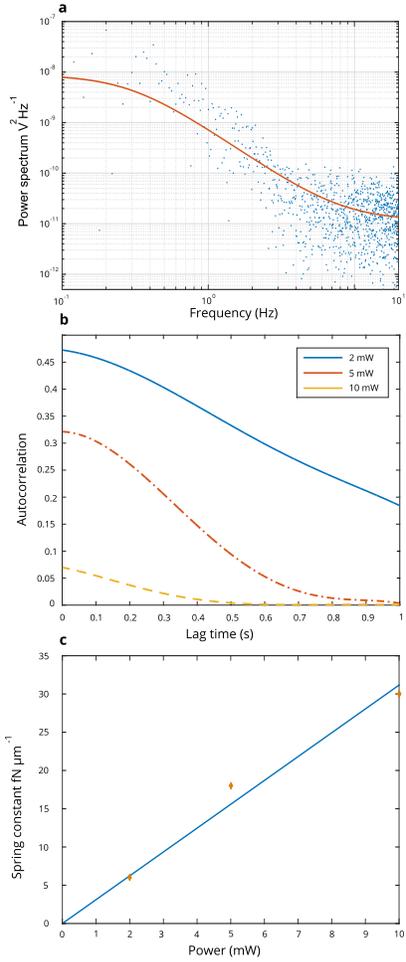}
\caption{\label{fig:fig6} (a) Power spectrum density of the tranmitted signal for 5 mW of trapping power. A corner frequency of 0.6 Hz is measured.(b) Autocorrelation signals at 2 mW, 5 mW and 10 mW. The observed decrease in the slope of the autocorrelation signal at different powers indicates a linear increase in trap strength with power as is expected. (c) Plot of the Spring constant as determined using the autocorrelation measurement vs. the power in the trapping beams. The subsequent plot is linear with respect to power as predicted.}
\end{figure}

Throughout the experiment we record the transmission through the STOF during trapping  and non-trapping events. Analyzing the power spectral density of this signal can be difficult when low trap frequencies are considered Fig. \ref{fig:fig6}(a). Additionally, our detected signal is coupled to the three non-degenerate trap stiffnesses and it is difficult to distinguish between single-particle and multiple-particle trapping; this adds more noise to our detected signal.  To overcome some of these issues, we opt for autocorrelation measurements which are analogous to the PSD, but do not suffer from some of the associated PSD measurement problems \cite{jones2015optical}. Data was taken in 50 s intervals at a sample rate of 2 kHz. The decay of the autocorrelation with respect to the delay time, given in Fig. \ref{fig:fig6}(b), is not quite exponential (as would be expected for a clean optical trap). The introduction of 'random' forces, can alter the lineshape \cite{roldan2014irreversibility}. We expect fluctuations in the slot walls due to external effects such as air currents etc., while immersed in the particle solution can alter the trapping force on the particles to be sources of these 'random' forces. Despite these contributions, the trap stiffnesses are seen to increase proportionally to the input powers, as determined by closest exponential fits to the data. The increase is directly proportional to the input laser power Fig. \ref{fig:fig6}(c) and give values which are of the same magnitude as the  expected theoretical values, Table \ref{tab:trapstiff}. We instead take the root mean square (RMS) value for the power to adjust for the losses along the fiber and use this power to calculate $k_x$, $k_z$, and $g_y$. The use of the RMS value adjusts the power for adiabatic losses in the taper region of the nanostructured fiber. We assume that the down taper and up taper sections of the STOF are symmetric, therefore losses accrued from both sections should be equal. Hence, we assume that $P_{slot}=\sqrt{T}$, where $P_{slot}$ is the power at the slot and $T$ is the transmission of the fiber. A similar argument can be used for the influence of the slot. Consideration of multiple particle interactions and surface-particle interactions such as the Faxen corrections would reduce this value further. We assume that the longitudinal trap stiffness, $k_z$, may have a smaller contribution to the measured trap strength since this trap corresponds to motion longitudinally along the fiber which would not result in significant noise contributions in our recorded signal. Additionally, the trap in the \textit{y}-direction does not have a restoring character which would alter its contributions to the noise spectra. This may point to the \textit{x} component of the trap being the primary contribution to the measured signal.

\section{Conclusions}
Micro-nanofibers have recently established themselves as very useful tools in several fields, including optical trapping and cold atom physics \cite{kumar2015multi}. The STOF used in the work reported here allowed us to further enhance the effectiveness of MNFs in optical trapping. As a trapping device, the STOF shows promise for particle sizes down to 200 nm with modest trapping powers, albeit with low trap stiffnesses. Trapping of 100 nm polystyrene particles has also been observed, but was not presented here. The flexibility of the fabrication process permits us to make structures with very high resolution, as well as providing a means for \textit{in-situ} scanning electron microscope (SEM) measurements of the device prior to use. This flexibility opens up many avenues of research as it facilitates arbitrary modification of the MNF waist. The unique trapping geometry which confines particles, and potentially atoms, within the slot leads us to believe that spectroscopic measurements are possible by passing probe beams of different wavelengths through the fiber while simultaneously recording the transmission or captured fluorescence at the output  pigtail. A substantial improvement to earlier work, such as the self-organization of atoms along nanophotonic waveguides \cite{chang2013self}, should also be possible since light coupling into the STOF is increased compared to for standard optical nanofibers.
This study serves as a step towards the realization of more complex applications involving the incorporation of different slot geometries as well as custom MNF Bragg gratings \cite{Kato15} to further enhance the fields of the STOF. Whether as a platform for studying optical binding effects, as an analytical tool, or as a trap for cold, neutral atoms, the STOF has many exciting applications which remain to be investigated.

\section*{Acknowledgments}
This work was supported by the Okinawa Institute of Science and Technology Graduate University. The authors would like to thank T. Sasaki and L. Szikszai for invaluable technical assistance and V. Brulis (Photon Design) for his insightful comments regarding the FDTD simulations.

\end{document}